\documentstyle[aps,prc]{revtex}
\begin{document}
\draft
\title{
Effects of $\beta_{6}$ deformation and low-lying vibrational bands on
 heavy-ion fusion reactions at sub-barrier energies
}
\author{Tamanna Rumin$^1$,
\thanks{E-mail address: rumin@nucl.phys.tohoku.ac.jp}
Kouichi Hagino$^2$, 
\thanks{E-mail address: hagino@phys.washington.edu}
and
Noboru Takigawa$^1$
\thanks{E-mail address: takigawa@nucl.phys.tohoku.ac.jp}
}
\address{$^1$ Department of Physics, Tohoku University, 
Sendai 980-8578, Japan \\
$^2$ Institute for Nuclear Theory, Department of Physics,
University of Washington, Seattle, WA 98195
}
\date{\today}

\maketitle

\begin{abstract}

We study fusion reactions of $^{16}$O with $^{154}$Sm, $^{186}$W and 
$^{238}$U at sub-barrier energies by a coupled-channels framework. 
We focus especially on the effects of $\beta_{6}$ deformation and 
low-lying vibrational excitations of the target nucleus. 
It is shown that the inclusion of $\beta_{6}$ deformation leads to a 
considerable improvement of the fit to the experimental data for all of 
these reactions. 
For the $^{154}$Sm and $^{238}$U targets, the octupole vibration 
significantly affects the fusion barrier distribution. 
The effect of $\beta$ band is negligible in all the three reactions, 
while the $\gamma$ band causes a non-negligible effect on the barrier 
distribution at energies above the main fusion barrier. 
We compare the optimum values of the deformation parameters obtained by 
fitting the fusion data with those obtained from inelastic scatterings 
and the ground state mass calculations. 
We show that the channel coupling of high multipolarity beyond the quadrupole 
coupling is dominated by the nuclear coupling and hence higher order 
Coulomb coupling does not much influence the optimum values of $\beta_4$ 
and $\beta_6$ parameters. 
We also discuss the effect of two neutron transfer reactions on the fusion 
of $^{16}$O with $^{238}$U.
\end{abstract}

\pacs{
25.70.Jj,  
21.60.Ev,    
24.10.Eq,     
23.20.Js  
}
\section{Introduction}

It is now well established that nuclear intrinsic motions significantly 
enhance the fusion cross-section in heavy-ion reactions at sub-barrier 
energies. 
Deformation effect is one of such prominent effects. The role of
static deformation in enhancing fusion cross section has long been 
recognized \cite{won73,vaz74} and has been experimentally demonstrated 
\cite{sto78,rei82,rei85}. 
Here the enhancement occurs because there is a distribution of barrier 
heights which can be thought of as resulting from different orientations 
of the deformed target nucleus. Any distribution of barriers around a 
single Coulomb barrier leads to enhancement of the fusion cross section 
at energies below the single barrier, because passes through the lower 
barriers are much more probable. 
Recently, high precision experimental data were obtained for 
the $^{16}$O+$^{186}$W, $^{154}$Sm fusion reactions and it was clearly 
demonstrated that sub-barrier fusion reactions strongly depend upon the 
nuclear hexadecapole deformation \cite{lem93,le95}. 
It was pointed out that the optimum values of the quadrupole and hexadecapole 
deformation parameters obtained from the analyses of such high precision 
fusion data are consistent with those obtained from the Coulomb excitation 
\cite{shaw74,Fis75} using similar radius parameter \cite{le93}.
In order to reach this conclusion, the authors of Refs. \cite{lem93,le95,le93} 
included up to the $\beta_4$ deformation in their analyses, neglecting 
higher order deformations such as $\beta_6$. 
On the other hand, the differential cross sections of inelastic alpha 
particle \cite{apo70} and proton \cite{palla83} scatterings from $^{154}$Sm 
and neutron scattering from $^{186}$W \cite{del82} show important effects 
of $\beta_6$ deformation of the target nuclei. 
The important role of $\beta_6$ deformation of the target nucleus has been 
shown also in the inelastic alpha and proton scatterings from $^{238}$U, 
which give optimum deformation parameters consistent to each other
\cite{her73,mos71,ron81,han82}. 
Although each reaction might be sensitive to different channels, it is not 
obvious whether the effects of $\beta_6$ deformation on fusion reactions 
are negligible. 
We also notice that there still remain noticeable discrepancies between the 
experimental and theoretical barrier distributions in Ref. \cite{le95} which 
included up to $\beta_4$ deformation. 
Although a better reproduction of the experimental data of the fusion 
excitation function as well as the fusion barrier distribution has been 
obtained by including the effects of low-lying two 2$^+$ and one 3$^-$ 
vibrations and a positive Q-value transfer channel, the role of higher 
order deformations has not yet been clarified. 
This motivated the present work, where we perform a detailed study of the 
effects of higher order deformation, especially the effects of $\beta_6$ 
deformation on heavy-ion fusion reactions.  

Besides clarifying the mechanism of heavy-ion fusion reactions, the study 
of the effects of higher order deformation is interesting to see the 
possibility of heavy-ion fusion reactions as a new powerful method of 
nuclear spectroscopy. This is another motivation of the present work. We 
therefore compare the optimum values of the deformation parameters obtained 
from the analyses of the fusion data with those from inelastic 
scatterings and the ground state mass calculations. In order to have 
reliable results, one has to take various channel coupling 
mechanisms into account which might cause effects of similar order. 
In this connection, we discuss in this paper the effects of vibrational 
excitations of deformed targets. We also pay attention to the role of 
higher order Coulomb coupling.  

For these purposes, we particularly study $^{16}$O + $^{154}$Sm, $^{186}$W, 
$^{238}$U fusion reactions, where high precision experimental data have 
been obtained \cite{le95,hin96}. 
We discuss the effects of channel coupling through the excitation function 
of the fusion cross section and the fusion barrier distribution \cite{row91}, 
which is defined as the second derivative of the product of the bombarding 
energy $E$ and the fusion cross section $\sigma_{F}$ with respect to $E$. 
Though the fusion barrier distribution has, strictly speaking, clear physical 
meaning only in the limit of sudden fusion, i.e. in the limit where the 
excitation energy of intrinsic excitations can be ignored, it has been shown 
that the concept still holds to a good approximation even for non-zero 
excitation energy \cite{HTB97}. 
This method has often been used to analyse high precision heavy-ion fusion 
data and is now well known to provide a very sensitive test of various 
channel coupling effects. 

The paper is organized as follows. 
In Sec. II, we present the results of the coupled-channels analysis which 
takes only the ground state rotational band into account. 
The main result is that $\beta_{6}$ deformation plays an important role in 
all the three reactions. 
For $^{186}$W, the magnitude as well as the sign of the $\beta_6$ are 
consistent with the results of ground state mass calculations and 
inelastic neutron scattering. 
However, the sign of $\beta_{6}$ for the Sm and U targets is predicted to be 
opposite to the result of other studies. 
In Sec. III, we examine the validity of the calculations used in Sec. II, 
which take full order for the nuclear coupling, while only the linear order 
for the Coulomb coupling into account. 
By performing coupled-channels calculations keeping up to the second order 
in the Coulomb coupling, we show that the high multipolarity couplings, 
i.e. the Y$_4$ and Y$_6$ couplings, are dominated by the nuclear coupling 
and hence non-linear Coulomb coupling does not almost alter the optimum 
values of $\beta_4$ and $\beta_6$.  
In Sec.IV we present the results of coupled-channels analyses for 
$^{16}$O + $^{154}$Sm and $^{238}$U fusion reactions which take octupole 
vibrations into account. 
These target nuclei have low-lying K=0$^-$ octupole bands, which are 
strongly excited by the Coulomb excitation. 
The E3 transition strength from the ground state to the 3$^-$ state is 
24 W.u. for $^{238}$U and 11 W.u. for $^{154}$Sm. 
We show that the octupole vibration significantly affects the fusion barrier 
distribution and modifies the optimum values of deformation parameters to 
fit the experimental data. 
Especially, it changes the sign of $\beta_6$ deformation to agree with the 
analyses of inelastic $\alpha $ and proton scatterings and the ground state 
mass calculations. 
Notice that there is no experimental evidence for the low-lying octupole 
K=0$^-$ band in $^{186}$W, suggesting its absence in this nucleus. 
All the three target nuclei have low-lying $\beta $ and $\gamma$ bands, 
whose interband E2 transition probabilities from the ground state 0$^+$ to 
the 2$^+$ member are: 1.0 and 4.4 W.u. in $^{154}$Sm, 8.9 and 1.4 W.u. 
in $^{186}$W and 3.0 and 1.5 W.u. in $^{238}$U for the $\gamma$ and 
$\beta $ bands, respectively. 
In Sec.V, we examine the effects of $\beta$ and $\gamma $ vibrations, and 
show that the effect of $\beta $ band is negligible, while the $\gamma $ 
band affects the fusion barrier distribution at high energies. 
Besides nuclear intrinsic excitations, nucleon transfer reactions between 
the colliding nuclei can enhance the low-energy fusion cross section. 
In Sec.VI, we study the effect of pair neutron transfer reactions on the 
$^{16}$O +$^{238}$U fusion reactions and discuss whether it explains the 
experimental fusion cross section which is systematically larger than 
the prediction of the coupled-channels calculations which ignore particle 
transfer reactions at low energies. 
We summarize the paper in Sec.VII. 
Appendix A is added to briefly explain the theoretical framework of the 
coupled-channels calculations we use. 
We also add Appendix B to show the structure of the higher order Coulomb 
coupling. 

\section{Effect of $\beta_6$ deformation}

In this section we present the results of coupled-channels calculations 
which take only the ground state rotational band of the target nucleus 
into account. 
We treat the projectile as inert, since its excitations can be well 
incorporated with a choice of the bare potential \cite{HTDHL97}. 
Instead of handling the full coupled-channels equations, we introduce 
the no-Coriolis approximation throughout this paper, and ignore the change 
of the centrifugal potential barrier due to the finite multipolarity of 
nuclear intrinsic excitations \cite{tak86,hag95}. 
This leads to considerable reduction of the dimension of the coupled-channels 
equations. 
We assume an axially symmetric deformation for the target nucleus and expand 
the radius up to the hexacontatetrapole deformation $\beta_6$. 
We introduce the sudden tunneling approximation, and set the excitation 
energy of the ground state K=0$^+$ rotational band to zero. 
Together with the no-Coriolis approximation, this leads to a set of 
decoupled eigen-channel problems, each of which corresponds to the fusion 
with a fixed orientation of the target nucleus. 
Accordingly, we first solve the Schr\"{o}dinger equation for a given 
orientation $\theta $ for each partial wave $J$ using the incoming wave 
boundary condition to obtain the tunneling probability $P_J(E,\theta)$. 
We then calculate the total tunneling probability $P_J(E)$  for each $J$ 
by taking average over all orientations as,    
\begin{eqnarray}
P_J(E)=\frac{1}{2}\int_0^{\pi}P_J(E,\theta)\sin\theta d\theta, 
\label{pene}
\end{eqnarray}
where the weight of the average has been determined by the ground state wave 
function of the deformed target, which is initially in the 0$^+$ state. 
The fusion cross section is then obtained by the standard partial wave sum. 
Once the fusion excitation function has been obtained, the fusion barrier 
distribution is calculated by the point difference formula of 
$\Delta E$=2MeV in the laboratory energy, whose value was employed in Refs. 
\cite{lem93,le95,le93} in analyzing the experimental data.  

We first determine the nuclear potential parameters for each target nucleus 
by fitting the fusion cross section larger than 200 mb by a potential 
model \cite{le95}. 
We then calculate the fusion cross section by switching on deformations of 
different multipolarity successively. 
At each step, we determine the values of the deformation parameters by 
$\chi^2$ fitting of the data of fusion excitation function and readjust the 
potential parameters. 
We use the values in Ref.\cite{le95} as the initial values. 
The results are shown in Fig.1. 
The left and right columns are the excitation function of the fusion 
cross section and the fusion barrier distribution, respectively. 
The top, center and bottom panels are for $^{154}$Sm, $^{186}$W and $^{238}$U 
targets, respectively. 
The dashed, the dotted and the solid lines represent the results of 
coupled-channels calculations including only $\beta_{2}$ deformation, 
$\beta_{4}$ in addition and $\beta_{6}$ as well, respectively. 
The dashed line for the $^{16}$O+$^{186}$W fusion reactions cannot be seen 
clearly in the fusion excitation function, because it overlaps with the 
solid line in the semi-logarithmic plot of the present scale. 
However, it is clearly separated from the other two lines in the fusion 
barrier distribution. 
This typically shows the high sensitivity of the fusion barrier distribution 
to different channel coupling effects. 
The deformation parameters obtained in the analysis are shown in the figure. 
Those obtained including only up to $\beta_4$ deformation for the $^{154}$Sm 
and $^{186}$W targets somewhat differ from those obtained in \cite{le95}, 
which are $\beta_2$=0.33, $\beta_4$=0.05 and $\beta_2$=0.31, 
$\beta_4$=$-$0.03, respectively. 
Since we use the same radius parameter as in Ref. \cite{le95}, these 
differences can probably be attributed to the different methods to 
calculate the fusion cross section in two works. 
We calculated it by numerically solving the Schr\"odinger equations, while  
Ref. \cite{le95} introduced the parabolic barrier approximation, which does 
not work at energies far below the barrier. 

The importance of the $\beta_{6}$ deformation can be clearly seen in the 
fusion barrier distribution for all cases, and in the fusion excitation 
function as well for the $^{154}$Sm and $^{238}$U targets. 
The agreement between the experimental data and the coupled-channels 
calculations concerning the fusion barrier distribution has been 
significantly improved above 56 and 66 MeV for $^{154}$Sm and $^{186}$W, 
respectively, by including $\beta_6$ deformation. 
The $\beta_6$ deformation removes a sharp peak at around 82 MeV in the 
fusion barrier distribution, which appears in the coupled-channels 
calculations without $\beta_6$ deformation, for the $^{238}$U target.

In Fig.2, we show the dependence of the fusion excitation function and the 
fusion barrier distribution on the $\beta_6$ parameter for the 
$^{16}$O + $^{154}$Sm, $^{186}$W reactions. 
The three lines in each figure have been calculated by using the same 
parameter sets as in Fig.1 (the solid lines), or by inverting the sign of 
the $\beta_6$ parameter (the dotted lines) and by setting it to be zero 
(the dashed lines). 
This figure also shows that the effect of $\beta_6$ deformation on the 
fusion cross section is not negligible. 

We compare in Table 1 the optimum values of the deformation parameters thus 
obtained with those obtained from the analyses of inelastic scatterings 
and the ground state mass calculations. 
The table also shows the radius parameter used in each analysis. 
For all three target nuclei, we observe noticeable discrepancies in the 
magnitudes of the deformation parameters among different studies 
(magnitude problem). 
This problem is, however, not so serious as it appears, because the 
discrepancies are largely due to the different choice of the radius parameter 
in each analysis. 
Since the strength of the channel coupling depends on the product of the 
deformation and radius parameters for the nuclear part and on 
$\beta_\lambda\times R_T^\lambda$ for the Coulomb part (see Eqs.(A.2--A.4)), 
physically important quantities are these products. Based on this idea, 
Table 1 shows the scaled deformation parameters as well, which have been 
calculated by $\beta_\lambda \times r_0/1.06$ (figures with a star) 
or by $\beta_\lambda \times (r_c/1.06)^{\lambda}$ (figures with two stars) 
from the original deformation parameters. 
We observe that the scaled deformation parameters from non-fusion studies are 
now much closer to the optimum deformation parameters from the fusion 
analysis. 
We cannot unfortunately rescale the deformation parameters for the neutron 
scattering from $^{186}$W, since the radius parameter is not given in 
\cite{del82}. 

We wish to especially remark that the sign and the magnitude of $\beta_6$ 
obtained from fusion analysis are consistent with those obtained from 
the ground state mass calculations for the $^{186}$W target. 
Our result is consistent also with the neutron scattering, though it gives 
only the upper bound of the magnitude. 
On the other hand, the predicted sign of $\beta_6$ is opposite to the results 
of other studies for $^{154}$Sm and $^{238}$U (sign problem). 
We show in Sec. IV that the effect of octupole vibration provides a 
possibility to cure this sign problem.
We note that the optimum deformation parameters of $^{154}$Sm and $^{238}$U 
obtained from many experiments of proton and $\alpha$ particle scatterings 
agree quite well to each other including the sign and magnitude of 
$\beta_6$, though there exist a few exceptions in the case of $^{154}$Sm. 

For reference, we compare in Fig.3 the experimental data and the results 
of coupled-channels calculations using deformation parameters from the fusion 
data (the solid line), the ground state mass calculations \cite{ato95} 
(the dotted line) and inelastic scatterings  (the dashed line). 
For the latter, we choose the results of inelastic alpha \cite{apo70}, 
neutron \cite{del82} and proton \cite{han82} scatterings for $^{154}$Sm, 
$^{186}$W and $^{238}$U, respectively.
In calculating the dotted and the dashed lines, the radius of the target 
nucleus $R_T$ has been adjusted such that $\beta_2 R_T$ equals that in the 
fusion calculations, where we chose $R_T$=1.06$\times$A$_T^{1/3}$ fm in 
order to have the same nuclear quadrupole coupling as for the solid line. 
Naturally, the parameter set obtained by the $\chi^2$ fit of the fusion data 
provides the best fit to the experimental data. 

\medskip

Table 1: Comparison of the optimum deformation parameters and nuclear 
radius parameter in various analyses. 

\begin{center}
\begin{tabular}{|l|l|l|l|l|l|l|l|l|l|l|}
\hline 
Nuclei &Methods &$r_0$ (fm) &$r_c$ (fm) & $\beta_2$ & $\beta_4$ 
& $\beta_6$ \\
\hline
$^{154}$Sm & $^{16}$O+$^{154}$Sm Fusion &1.06&....&0.322 & 0.027 & 0.027\\
& Mass Calculation\cite{ato95} &1.16&....&0.27 & 0.113 & $-$0.005 \\
          &                    &    &  &0.295$\ast$
&0.124$\ast$&$-$0.005$\ast$\\
&$\alpha$ Scattering \cite{apo70}&1.492& &0.225$\pm$0.005 &0.050$\pm$0.005 
&$-$0.015$\pm$0.010\\
          &                    &    &  &0.317$\ast$&0.070$\ast$
&$-$0.021$\ast$\\
&proton Scattering \cite{palla83}& &....&0.285 & 0.051 & $-$0.015\\
\hline
$^{186}$W  & $^{16}$O+$^{186}$W Fusion &1.06&.... &0.285 & $-$0.031 & 0.027\\
& Mass Calculation \cite{ato95}&1.16&.... &0.23 & $-$0.107 & 0.02\\
          &                    &    &  &0.25$\ast$&$-$0.117$\ast$&0.022$\ast$\\
 &neutron Scattering\cite{del82}&....&....&0.203$\pm$0.006
&$-$0.057$\pm$0.006& $<\vert$$-$0.04$\vert$\\
\hline
$^{238}$U & $^{16}$O+$^{238}$U Fusion &1.06&....&0.289 & 0.01 & 0.044\\
& Mass Calculation \cite{ato95}&1.16& ....&0.215 & 0.093 & $-$0.015\\
          &                    &    &  &0.235$\ast$&0.102$\ast$
&$-$0.016$\ast$\\
 &$\alpha$ Scattering \cite{her73}& 1.2&....&0.22$\pm$0.01& 0.06$\pm$0.01 
& $-$0.012$\pm$0.01\\
          &         &    &  &0.25$\ast$&0.068$\ast$&$-$0.014$\ast$\\
&proton Scattering \cite{han82}& ....&1.25&0.225$\pm$0.005 & 0.045$\pm$0.005 
& $-$0.015$\pm$0.003\\
          &         &    &  &0.313$\ast\ast$&0.087$\ast\ast$
&$-$0.040$\ast\ast$\\
\hline
\end{tabular}
\end{center} 

\section{Higher order Coulomb coupling}

The results in Sec. II have been obtained by treating the Coulomb coupling 
in the linear order and the nuclear coupling in full order. 
Though this approximation is often used in literatures, it is worth checking 
the validity, especially in discussing the role of higher order deformations. 
One would guess that this approximation breaks down when the charge product 
of the projectile and target gets large. 
As examples, we performed coupled-channels calculations for the 
$^{32}$S + $^{168}$Er and $^{16}$O + $^{154}$Sm fusion reactions by including 
the second order Coulomb coupling and by assuming only the quadrupole 
coupling. 
We found that the second order Coulomb coupling noticeably modifies the 
fusion barrier distribution. 
Naturally, the modification is more significant for the former reaction. 
An important issue in the context of the present paper is whether the higher 
order Coulomb coupling significantly changes the optimum values of higher 
order deformation parameters that reproduce the experimental data of fusion 
cross section. 
In this connection, we show in Appendix B the higher order terms in the 
Coulomb interaction up to the order of $\beta_6$, i.e. $\beta_4\times\beta_2$. 
Although $\beta_2^3$ would contribute in the same order, we do not show it, 
since it is very tedious to evaluate it and also its effects are negligible 
as we argue below. 
Note that other terms, like $\beta_4^2$ and $\beta_2\times\beta_6$, are 
higher order contributions, which are the same order of $\beta_8$ or higher, 
and are not shown. 
Eq.(B.1) indicates that the optimum values of $\beta_4$ and $\beta_6$ 
parameters will be considerably altered by the non-linear coupling if the 
Coulomb coupling significantly contributes to the higher multipolarity, 
i.e. Y$_4$ and Y$_6$, couplings. 

In order to examine the situation, we compare in Fig.4 the fusion barrier 
distribution calculated in four different ways. 
For simplicity, all the calculations have been performed by treating both 
the nuclear and Coulomb couplings in linear order and by expanding up to the 
Y$_6$ term. 
The solid line is the fusion barrier distribution obtained by keeping both the 
nuclear and Coulomb couplings as they are. 
The dashed line has been obtained by discarding the nuclear Y$_4$ coupling 
term. 
We observe a significant change of the fusion barrier distribution. 
We performed additional two calculations, where only the Coulomb or both the 
nuclear and Coulomb Y$_4$ couplings are discarded. 
Their results are almost the same as the solid and the dashed lines, 
respectively. 
These results indicate that the Y$_4$ coupling is far dominated by the 
nuclear coupling. 
We checked that a similar situation holds also for the Y$_6$ coupling. 
We thus conjecture that the main effect of the Coulomb coupling resides in 
the Y$_2$ coupling, and one can determine to a good approximation the optimum 
$\beta_4$ and $\beta_6$ parameters through the coupled-channels calculations 
using the linear Coulomb coupling. 
In the following analyses, we thus treat the Coulomb coupling in the linear 
order. 
Keep, however, in mind that the optimum $\beta_2$ value could be noticeably 
affected depending on whether one uses the linear or higher order Coulomb 
coupling.  

\section{Effect of octupole vibration}

We now study the effects of octupole vibration on the $^{16}$O + $^{154}$Sm 
and $^{16}$O + $^{238}$U fusion reactions. 
As already mentioned in the introduction, there exist low-lying K=0$^-$ 
octupole bands in $^{154}$Sm and $^{238}$U, which are strongly excited by 
the Coulomb excitation through the E3 transition. 
We take into account their effects on fusion by solving coupled-channels 
equations for each orientation of the deformed target. 
We call this procedure the $\theta$-scheme. 
We confirmed that the results are almost the same as those obtained by 
treating the rotational excitations not by the $\theta$-scheme, but by 
specifying each excited level by its spin, and by solving coupled-channels 
equations with a larger dimension which include both the K=0$^{+}$ ground 
state and K=0$^{-}$ octupole bands \cite{Tamanna}. 
Similarly to the rotational coupling, we treat the nuclear part of the 
octupole coupling in full order, while the Coulomb part in the linear order. 
The amplitude of the zero point motion of the octupole vibration, which 
governs the strength of the channel-coupling, is determined from the 
experimental value of the reduced transition probability B(E3$_o\uparrow$) 
\cite{nds87,nds88} from the ground state to the 3$^-$ state of the K=0$^-$ 
octupole vibrational band following    
\begin{eqnarray} 
\alpha_0^o=\displaystyle{\frac{(\frac{4\pi}{3ZR_T^3})
\sqrt{B(E3_o\uparrow)/e^2}}{\Big[1+\frac{1}{3}\sqrt{\frac{5}{\pi}}\beta_2+
\frac{5}{22}\sqrt{\frac{9}{\pi}}\beta_4+\frac{125}{13\times 11
\times 3}\sqrt{\frac{13}{\pi}}\beta_6\Big]}}.
\label{alpha0}
\end{eqnarray}

The optimum set of deformation parameters as well as the potential parameters 
are readjusted by the $\chi^2$ fitting after including the $3^-$ vibrational 
state. 
The results are shown in Fig.5 by solid lines in comparison with the 
experimental data and the previous calculations which include only the 
ground state rotational band (the dashed lines). 
The effects of the octupole vibration are visible especially in the fusion 
barrier distribution. 
The resultant optimum deformation parameters are 
$\beta_2$=0.314, $\beta_4$=0.011 and $\beta_6$=$-$0.016 for $^{154}$Sm and 
$\beta_2$=0.279, $\beta_4$=0.0007 and $\beta_6$=$-$0.024 for $^{238}$U. 
An interesting result is that the sign problem of $\beta_6$ parameter has 
been resolved for both $^{154}$Sm and $^{238}$U nuclei, although the optimum 
values of $\beta_4$ become too small, especially in $^{238}$U, compared with 
the other analyses. 

\section{Simultaneous effects of octupole, $\beta$ and $\gamma$ vibrations}

We now add the effects of the $\beta$ and $\gamma$ vibrations. 
We treat all the vibrational excitations by a coupled-channels framework 
by keeping their finite excitation energies and using the linear coupling 
approximation not only for the Coulomb but also for the nuclear parts. 
The rotational coupling is treated in the same way as in Sec. II. 

The amplitudes of the zero point motion of the $\beta$ and $\gamma$ vibrations 
are determined from the experimental values of the reduced transition 
probability B(E2$\uparrow$) \cite{fire88} from the ground state to the 
2$^+$ state of the $\beta$ band and to the band head of the $\gamma $ band. 
The formulae we use are,
\begin{eqnarray}
\alpha_0{^\beta}=\displaystyle{\frac{\sqrt{B(E2_{\beta}\uparrow)/e^2}}
{(\frac{3ZR_T^2}{4\pi})\Big(1+\frac{4}{7}\sqrt{\frac{5}{\pi}}\beta_2\Big)}},
\hspace{0.5cm}
\alpha_0{^\gamma}=\displaystyle{\frac{\sqrt{B(E2_{\gamma}\uparrow)/2e^2}}
{(\frac{3ZR_T^2}{4\pi})\Big(1-\frac{4}{7}\sqrt{\frac{5}{\pi}}\beta_2\Big)}}.
\label{amplbg}
\end{eqnarray}
Table 2 collects the experimental transition probabilities and the values 
of the zero point motion amplitudes for the $\beta$ and $\gamma$ vibrations 
together with those for the octupole vibration. 
It includes also $\beta_2$ values. 
They have been extracted in Sec. II and used to determine $\alpha_0^{\beta}$, 
$\alpha_0^{\gamma}$ and $\alpha_0^{o}$ following Eqs. \ref{amplbg} and 
\ref{alpha0}. 

\medskip

Table 2 : The zero point motion amplitude of the octupole, $\beta$ and 
$\gamma$ vibrations.

\begin{center}
\begin{tabular}{|l|l|l|l|l|l|l|l|l|l|l|l|l|}
\hline 
Nuclei&$B(E3_o\uparrow) (e^2b^3) $&$B(E2_{\beta}\uparrow) (e^2b^2)$
&$B(E2_{\gamma}\uparrow) (e^2b^2) $&$\beta_2$  &$\alpha_0^o$ 
& $\alpha_0^{\beta}$&$\alpha_0^{\gamma}$\\
\hline
$^{154}$Sm& 0.100\cite{nds87}&0.023\cite{fire88}& 0.069\cite{fire88} & 0.322 & 0.103 & 0.026 & 0.051\\
$^{186}$W & ..... & 0.009\cite{fire88}& 0.150\cite{fire88}& 0.285 & ...... & 0.012 & 0.054\\
$^{238}$U & 0.575\cite{nds88} & 0.0656\cite{fire88} & 0.131\cite{fire88} & 0.289 & 0.109 & 0.0224 & 0.034\\
\hline
\end{tabular}
\end{center} 

\medskip

Since the coupling to the $\gamma$ band depends on the second Euler angle 
$\phi$, we first solve the coupled-channels equations for a given set of 
($\theta$,$\phi$) parameters. 
The fusion cross section for each partial wave $J$ is then calculated by 
taking average over both $\theta$ and $\phi$, 
\begin{eqnarray}
P_J(E)=\frac{1}{4\pi}\int_0^{\pi}\sin\theta d\theta \int_0^{2\pi} d\phi 
P_J(E,\theta,\phi).
\label{pene2}
\end{eqnarray}
The integrations are performed by Gauss quadrature. 
Since the numerical computation is quite heavy, we have not optimized the 
deformation parameters, but fixed them to those values used to obtain 
the solid line in Fig.1 for $^{16}$O+$^{186}$W and those in Fig.5 for 
$^{16}$O+$^{154}$Sm and $^{16}$O+$^{238}$U fusion reactions.  

The results are shown in Fig.6. Similarly to Fig.1, the 
fusion excitation function and the fusion barrier distribution are shown 
on the left and right sides, respectively. The solid line for the 
$^{16}$O+$^{186}$W fusion reaction was obtained by including the effect 
of only the ground state rotational band, and is the same as the solid 
line in Fig.1, while the solid line for the $^{16}$O+$^{154}$Sm and 
$^{16}$O+$^{238}$U fusion reactions was obtained by adding the octupole 
vibration. 
It is slightly different from the solid line in Fig.5 because of the 
different treatment of the nuclear coupling.

The effect of the $\beta$ band is very small for all three systems and 
invisible in the scale of Fig.6. 
The dotted line was obtained by adding the effect of $\gamma$ band. 
Its effect is less important than that caused by the $\beta_4$ and $\beta_6$ 
deformations, but is noticeable in the fusion barrier distribution. 
An interesting thing is that the $\gamma$ band does not affect the fusion 
excitation function at low energies. 
Consequently, its effect cannot be seen clearly in the fusion excitation 
function and concentrates in relatively high energy region in the fusion 
barrier distribution. 
This contrasts with higher order deformations, which affect the fusion 
excitation fusion at low energies as well, and hence the fusion barrier 
distribution over all energy region. 

\section{Effect of pair neutron transfer}

Before we close the paper, we would like to comment on possible effects 
of pair neutron transfer channel on the fusion reactions. 
Refs. \cite{bro85,esb89} claim that positive Q-value pair neutron transfer 
channels explain the isotope effects,  seen for example in 
$^{58}$Ni + $^{58}$Ni, $^{58}$Ni + $^{64}$Ni and $^{64}$Ni + $^{64}$Ni 
fusion reactions by enhancing the fusion cross section at low energies in 
$^{58}$Ni + $^{64}$Ni collision. 
Similarly, by studying $^{28}$Si + $^{68}$Zn scattering at sub-barrier 
energies, Ref. \cite{kat97} claims that the coupling of the positive Q-value 
two neutron transfer channel significantly enhances the fusion cross section. 

Among the three reactions which we discuss in this paper, only the 
$^{16}$O + $^{238}$U  has a two neutron transfer channel whose Q-value is 
positive, the Q-value for the two neutron pick-up reaction from 
$^{16}$O + $^{238}$U to $^{18}$O + $^{236}$U being 0.826 MeV in the ground 
state channel. 
This transfer channel might resolve the discrepancy between the experimental 
data and the coupled-channels calculations in the fusion excitation function 
at low energies for the $^{16}$O + $^{238}$U fusion reaction (see Fig.6). 
In order to see this possibility, we study here the effects of this transfer 
channel following the prescription in Ref.\cite{bro85}, where the transfer 
reaction is treated in the same way as a vibrational excitation in the 
coupled-channels formalism. The form factor of the transfer reaction is 
assumed to be 
\begin{eqnarray}
F_{trans}(R,\theta) = -\sigma_t\frac{dV_N(R,\theta)}{dR}, 
\end{eqnarray}
where $\sigma_t$ is the strength parameter of the transfer reaction, 
$V_N(R,\theta)$  is the deformed ion-ion potential. 
This form factor is slightly simplified from that in Refs. \cite{lan88,das86} 
by ignoring a small correction term. 
We determine the strength parameter by fitting the excitation function of 
the fusion cross sections. 
The optimum set of deformation parameters are readjusted by the $\chi^2$ 
fitting to the experimental data after including pair neutron transfer. 

The results are shown in Fig.7. The solid line includes the effect of the 
transfer reaction, while the dashed line takes only the rotational 
excitation into account. 
We left out the effects of vibrational coupling in these calculations. 
We see that the transfer channel significantly enhances the fusion cross 
section at low energies. 
The optimum deformation parameters in this analysis are 
$\beta_2$=0.299, $\beta_4 $ = 0.002 and $\beta_6$=0.034. 
Unfortunately, the dimension of the coupled-channels calculations becomes 
too large to simultaneously take both effects of transfer reaction and 
vibrational excitations into account. 
In the summary section, we also mention the possible importance of single 
nucleon transfer reactions.

\section{Summary}        

We studied the effects of $\beta_{6}$ deformation on heavy ion fusion 
reactions at energies near and below the Coulomb barrier by analyzing the 
excitation function of the fusion cross section and fusion barrier 
distribution for $^{16}$O+$^{154}$Sm, $^{186}$W and $^{238}$U reactions. 
Coupled-channels equations have been solved by taking the rotational 
excitation, and the octupole, $\beta$ and $\gamma$ vibrations of the 
target nuclei into account stepwise. 
The calculations which took only the ground state rotational band into 
account showed that $\beta_{6}$ deformation is important for all three 
reactions. 
The optimum value of $\beta_{6}$ well agrees with that obtained from the 
ground state mass calculations and inelastic neutron scattering for $^{186}$W. 
On the other hand, the sign of $\beta_{6}$ is inconsistent with that from the 
analyses of inelastic alpha and proton scatterings and the ground state mass 
calculations for $^{154}$Sm and $^{238}$U targets.  
We examined the validity of the linear approximation we took for the Coulomb 
coupling and gave a reasoning to conjecture that it is good enough to 
determine the optimum values of higher order deformation, i.e. 
$\beta_4$ and $\beta_6$, parameters.  

We have then shown that the coupling to the low-lying octupole vibration 
significantly affects the fusion barrier distribution in the 
$^{16}$O+$^{154}$Sm, $^{238}$U reactions. 
Interestingly, it changed the sign of the optimum $\beta_6$ to agree with 
that suggested from non-fusion analyses. 
The $\beta$ and $\gamma$ vibrations are then also taken into account. 
We found that the $\beta$ vibration introduces only negligible effect, while 
the $\gamma$ vibration changes the fusion barrier distribution by a 
noticeable amount, though the change is less than that due to higher order, 
i.e. $\beta_4$ and $\beta_6$, deformations. 
An interesting feature is that the $\gamma$ band does not affect the fusion 
excitation function at low energies, and hence its effect mostly appears in 
the fusion barrier distribution in relatively high energy region. 
This contrasts with higher order deformations, which influence the fusion 
cross section at low energies as well, and hence the fusion barrier 
distribution over all energy region. 
We left $\chi^2$ fitting to optimize the deformation parameters to a future 
work because of the computational heaviness.

A problem with the $^{16}$O+$^{238}$U fusion reaction is that the 
coupled-channels calculations which include only rotational and vibrational 
excitations cannot reproduce large experimental fusion cross section at 
low energies. 
We showed in sect.VII that two neutron transfer reaction enhances the fusion 
cross section at low energies. 
One will, however, need to study the effects of single nucleon transfer 
reactions as well in order to reach a comprehensive understanding of this 
reaction and to draw a conclusive conclusion. 
In this connection, it is interesting to notice that larger experimental 
cross sections for one nucleon transfer reactions than those for two nucleon 
transfer reactions at low energies have been reported for several systems 
\cite{Robert93,Corradi93,mahon97}.

Finally, we wish to make some comments on the limitations 
of our theoretical framework. We assumed a simple Coulomb 
interaction given by Eq.(\ref{coul}), which has a few shortcomings. 
The first is that the bare Coulomb interaction is identified with 
the Coulomb interaction between two point charges instead 
of the Coulomb potential for a uniformly charged extended 
object, which is often used for heavy-ion collisions. 
The second is that the same Coulomb coupling form factor, 
which is valid only in the region where there is no 
overlap between the projectile and target nuclei, 
is used over all separation distance. Furthermore, 
the Coulomb and the nuclear deformation parameters are 
assumed to be the same. 
The first two problems are related to each other, 
and might cause non-negligible effects, 
since the turning point at the inner side of the 
potential barrier is located sometimes inside the standard 
choice of the so called Coulomb radius, $R_C=1.2(A_P^{1/3}+
A_T^{1/3})$ fm, or even the sum of the radii of the 
projectile and target nuclei. 
The analyses which allow different values for the Coulomb and the 
nuclear deformation parameters carry one of the important advantages of 
heavy-ion fusion reactions compared with the other analyses, say 
neutron scattering. Such analyses  
will explore the difference between the 
charge and matter distributions, and will be very interesting 
also in connection with the study of the structure of 
exotic unstable nuclei, which is one of the current interests of 
nuclear physics. 
We will discuss in detail the effects 
of the improvements of theoretical analyses in these three respects  
in a separate paper \cite{tti99}. 
Here we simply wish to mention that the first effect, i.e. the 
difference of the Coulomb interaction between two point charges 
from the Coulomb potential for a uniform charge distribution, 
can be mimiced to a large extent by renormalizing the bare potential. 

\section*{Acknowledgments}

We thank D.M. Brink, A.B. Balantekin, A. Ono, M. Ueda, N. Ihara 
and K. Yoshizaki for useful discussions. 
This research was supported by the Monbusho Scholarship and 
the International Scientific Research Program: Joint Research: contract 
number 09044051 from the Japanese Ministry of Education, Science and Culture. 

\appendix
\section{Brief description of the coupled-channels formalism}

In this appendix we briefly explain the coupled-channels formalism which 
we used. 
We present here the case, where all the $\beta$, $\gamma$ and octupole 
vibrations are taken into account. The total Hamiltonian reads
\begin{eqnarray}
 H =T+H_{int}(\xi)+V({\mbox{\boldmath $R$}},\xi),
\end{eqnarray}
where $T$ is the kinetic energy of the relative motion between the projectile 
and target, $H_{int}(\xi)$ the Hamiltonian of the intrinsic motions of 
the colliding nuclei, whose coordinates are denoted by $\xi$, and 
$V({\mbox{\boldmath $R$}},\xi)$ the interaction Hamiltonian which depends 
on the coordinates of both the relative motion, $\bf R$, and nuclear 
intrinsic motions. 

We use the geometrical collective model for nuclear intrinsic motions. 
The variables $\xi $ are then the static as well as dynamic deformation 
parameters specifying the radius of the target nucleus as,
\begin{eqnarray}
R(\theta,\phi,a)=R_{T}\Big[1+\sum_{\lambda}\beta_{\lambda}Y_{\lambda 0}
(\theta)+a^\prime_{2 0}Y_{2 0}(\theta)+a^\prime_{2 2}[Y_{2 2}(\theta,\phi)
+Y_{2-2}(\theta,\phi)]+a_{3 0}Y_{3 0}(\theta)\Big].
\label{radius}
\end{eqnarray}
In writing Eq.(\ref{radius}) we chose the rotating coordinate frame, where 
the z-axis is taken to be parallel to the coordinate of the relative 
motion $\bf R$ \cite{hag95}. 
$\theta$ and $\phi$ are Euler angles which define the orientation of the 
principal axes of the deformed target in this frame. 
$\beta_\lambda$, $\lambda$ being 2, 4 and 6, are the static deformation 
parameters. $a$ means $a^\prime_{20}$, $a^\prime_{22}$ and $a_{30}$ which are 
the dynamical deformation parameters describing the $\beta$, $\gamma$ 
and octupole K=0$^-$ vibrations, respectively.

The interaction Hamiltonian $V({\mbox{\boldmath $R$}},\xi)$ consists of the 
nuclear and Coulomb parts. We assume the former to be
\begin{eqnarray}
V_N(R,\theta,\phi,a)=\frac{-V_0}{1+\exp[(R-R_P-R(\theta,\phi,a))/a_0]}.
\end{eqnarray}
It contains both the bare potential and the coupling Hamiltonian. When we 
treat the channel-coupling in the perturbation theory, say, of first or 
second order, we expand $V_N(R,\theta,\phi,a)$ with the relevant deformation 
parameters. The actual procedure of the full order coupled-channels 
calculations is explained in Refs. \cite{hagino97,HRK99}. 

We assume a simple Coulomb interaction by ignoring the change 
of the analytic expressions of the bare Coulomb interaction 
and the Coulomb coupling form factor 
depending on the relative magnitude between the distance R and  
either the sum of the charge radii of the projectile and target 
or the absolute value of their difference. The formula we take reads 
up to the leading order of the dynamical variables as
\begin{eqnarray}
V_C(R,\theta,\phi,a)&=&\frac{Z_PZ_Te^2}{R}+\sum_{\lambda}
\frac{3Z_PZ_Te^2}{2\lambda+1}\frac{R_T^\lambda}{R^{\lambda+1}}\Big
[\beta_{\lambda}Y_{\lambda 0}(\theta)+
a^\prime_{\lambda 0}Y_{\lambda 0}(\theta)\delta_{\lambda,2}\nonumber\\
&&+\delta_{\lambda,2}a^\prime_{\lambda 2} \Big(Y_{\lambda 2}(\theta,\phi)
+Y_{\lambda -2}(\theta,\phi)\Big)\Big]+\frac{3Z_PZ_Te^2}{7} a_{3 0}Y_{30}
(\theta)\frac{R_T^3}{R^4}.
\label{coul}
\end{eqnarray}
We assume the same charge radius and deformation parameters 
as those for the nuclear part for the target nucleus. 

The Hamiltonian for the intrinsic motions consists of four parts, 
\begin{eqnarray}
H_{int}(\xi)=H_{rot}+H_{\beta}+H_{\gamma}+H_{o}.
\end{eqnarray}
They describe the rotational and $\beta$, $\gamma$ and octupole vibrational 
excitations. 
Their explicit forms and the corresponding eigenfunctions and eigenvalues can 
be found in Ref.\cite{eis87}

We introduce two basic approximations. 
The one is the no-Coriolis approximation and the other the sudden tunneling 
approximation, i.e. degenerate spectrum approximation, for the rotational 
motion. 
The latter corresponds to setting $H_{rot}$ to be zero. 
In these approximations, coupled-channels equations are solved for each 
given set of ($J,\theta$,$\phi$), $J$ being the initial angular momentum of 
the relative motion, by expanding the wave function as
\begin{eqnarray}
\Psi_{J\theta\phi}(R,\xi_v)=\sum_n \frac{\chi^{J\theta\phi}_n(R)}{R}
\Phi_n(\xi_v),
\end{eqnarray}
where $\xi_v$ represent the coordinates of the $\beta$, $\gamma$ and 
octupole vibrations, and $n$ is the abbreviation of a set of corresponding 
quantum numbers $(n_\beta, n_\gamma,n_o)$ \cite{eis87}. 
Note that the angular part of the wave function for the relative motion 
is simply a constant in the rotating frame approximation. 
We consider only 0 or 1 for all the vibrational quantum numbers. 
The coupled-channels equations read
\begin{small}
\begin{eqnarray}
\Big[-\frac{\hbar^2}{2\mu}\frac{d^2}{dR^2}+\frac{J(J+1)\hbar^2}{2\mu R^2}+\epsilon_m+V(R,\theta,a=0)-E\Big]\chi^{J\theta\phi}_m(R)
=-\sum_nV_{mn}(R,\theta,\phi)\chi^{J\theta\phi}_n(R),
\end{eqnarray}
\end{small}
where $\epsilon_m$ is the eigenvalue of the vibration excitations 
corresponding to the eigenstate $\Phi_m(\xi_v)$. 
We represented the total interaction by separating it into the diagonal 
$V(R,\theta,a=0)$ and the explicit coupling $V_{mn}$ terms with respect to 
the vibrational excitations. The latter have been evaluated using the wave 
functions for vibrational motions given in Ref. \cite{eis87}. 

We solve the coupled-channels equations by imposing the incoming wave 
boundary condition at the position of the s-wave potential minimum, and 
determine the fusion probability by evaluating the incoming flux in each 
channel at that position. 
Once the fusion probability is obtained in this way for a given set of 
($J, \theta$,$\phi$), the total fusion probability for that partial wave is 
calculated by taking average over the orientation ($\theta$,$\phi$) as given 
by Eq.(\ref{pene2}). 
The fusion cross section is then calculated by the usual partial wave sum.

\section{Higher order Coulomb coupling}

Here we present the explicit form of the Coulomb coupling up to the 
Y$_6$ term when the second order coupling terms are included. 
Only the major terms are explicitly shown for the second order coupling. 

\begin{small} 
\begin{eqnarray}
V_C(R,\theta)&=&\frac{Z_PZ_Te^2}{R}+\frac{3}{5}\left(\beta_2
+\beta_2^2\frac{2}{7}\sqrt{\frac{5}{\pi}}
+\beta_2\beta_4\frac{4}{7}\sqrt{\frac{9}{\pi}}
\right)Z_PZ_Te^2\frac{R_T^2}{R^3}Y_{20}(\theta)\nonumber\\
&+&\frac{3}{9}\left(\beta_4+\beta_2^2\frac{9}{7}\sqrt{\frac{1}{\pi}}
+\beta_2\beta_4\frac{300}{77}\sqrt{\frac{5}{\pi}}
\right)Z_PZ_Te^2\frac{R_T^4}{R^5}Y_{40}(\theta)\nonumber\\
&+&\frac{3}{13}\left(\beta_6+\beta_2\beta_4\frac{20}{143}
\sqrt{\frac{45\times 13}{\pi}}
\right) Z_PZ_Te^2\frac{R_T^6}{R^7}Y_{60}(\theta)
\end{eqnarray}
\end{small}


\newpage
\begin{center}
{\Large Figure Captions}
\end{center}

\noindent
{\large Fig.1}\\
Comparison of theoretical (a) excitation functions of the fusion cross 
section and (b) fusion barrier distributions with experimental data 
\cite{le95,hin96}. 
Only ground state rotational excitations are taken into account in the 
theoretical calculations.  
The dashed lines represent the optimum fits when only quadrupole deformation 
is included. 
The dotted curves include hexadecapole deformation, while the solid lines 
show the final fits including hexacontatetrapole deformation.\\

\noindent
{\large Fig.2}\\
Sensitivity of (a) the excitation function of the fusion cross section, and
 (b) the fusion barrier distribution to the hexacontatetrapole deformation.
The solid lines were obtained by using the optimum $\beta_6$ parameter. 
The dotted lines were obtained by inverting the sign of $\beta_6$, 
while the dashed lines represent the results when the hexacontatetrapole
 deformation is set equal to zero.\\

\noindent
{\large Fig.3}\\
Comparison of theoretical (a) excitation functions and (b) fusion barrier
 distributions calculated by using deformation paramaters from different 
analyses. The data are taken from Refs. \cite{le95,hin96}. 
The results using the 
optimum deformation parameters from the analysis of the fusion data are 
shown by the solid lines. The dashed lines use the deformation parameters 
obtained from inelastic scattering \cite{apo70,del82,han82}, while the 
dotted curves those from the nuclear ground state mass calculations 
\cite{ato95}. All the calculations include hexacontatetrapole deformation.\\

\noindent
{\large Fig.4}\\
Study of the relative importance between the nuclear and the Coulomb $Y_4$ 
couplings. 
The solid line has been calculated by including both nuclear and Coulomb 
Y$_4$ couplings, while the dashed line by ignoring the nuclear Y$_4$ coupling. 
The results, where only the Coulomb Y$_4$ coupling and both the nuclear and 
Coulomb Y$_4$ cuplings have been discarded are almost the same as the solid 
and the dashed lines, respectively.\\ 

\noindent
{\large Fig.5}\\
 Effects of octupole vibration. The dashed lines represent the optimum fits 
when only the ground state rotation band is taken into account, while the 
solid lines show the optimum fits when the coupling to the K=0$^-$ octupole 
vibration is added.\\

\noindent
{\large Fig 6}\\
The same as Fig.1, but when octupole, $\beta $ and $\gamma $ vibrations are 
taken into account. The solid line for the $^{186}$W target has been 
obtained by including only the ground state rotational band, while that for 
the $^{154}$Sm and $^{238}$U targets by adding the octupole vibration. 
The change due to the beta band is invisible. 
The dotted line has been obtained by adding the $\gamma$ band.\\  

\noindent
{\large Fig 7}\\
Effect of pair neutron transfer reaction on the $^{16}$O+$^{238}$U fusion 
reactions. 
The dashed line takes only the ground state rotational excitation into 
account, while the solid line the pair neutron transfer reaction in addition.

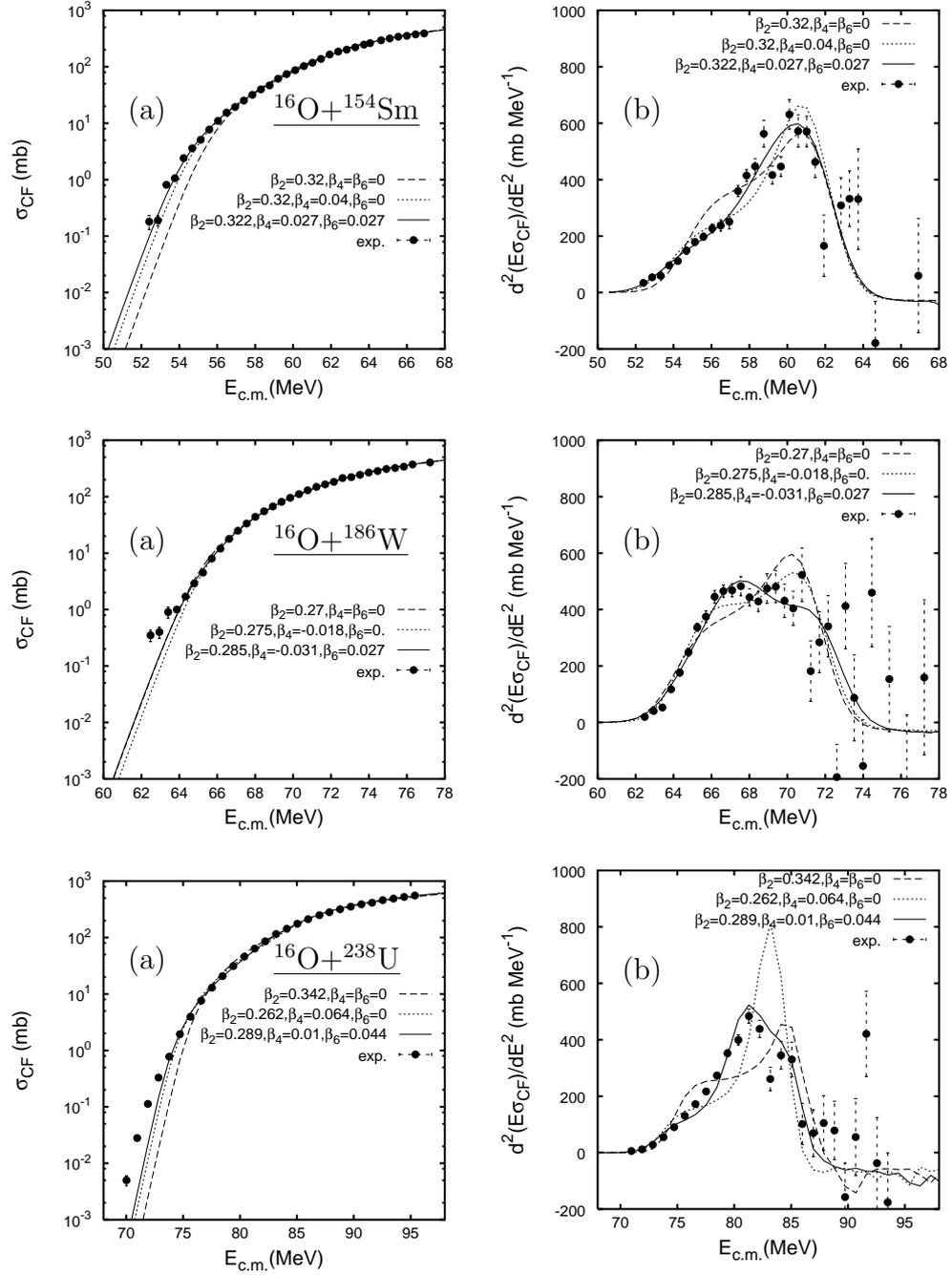
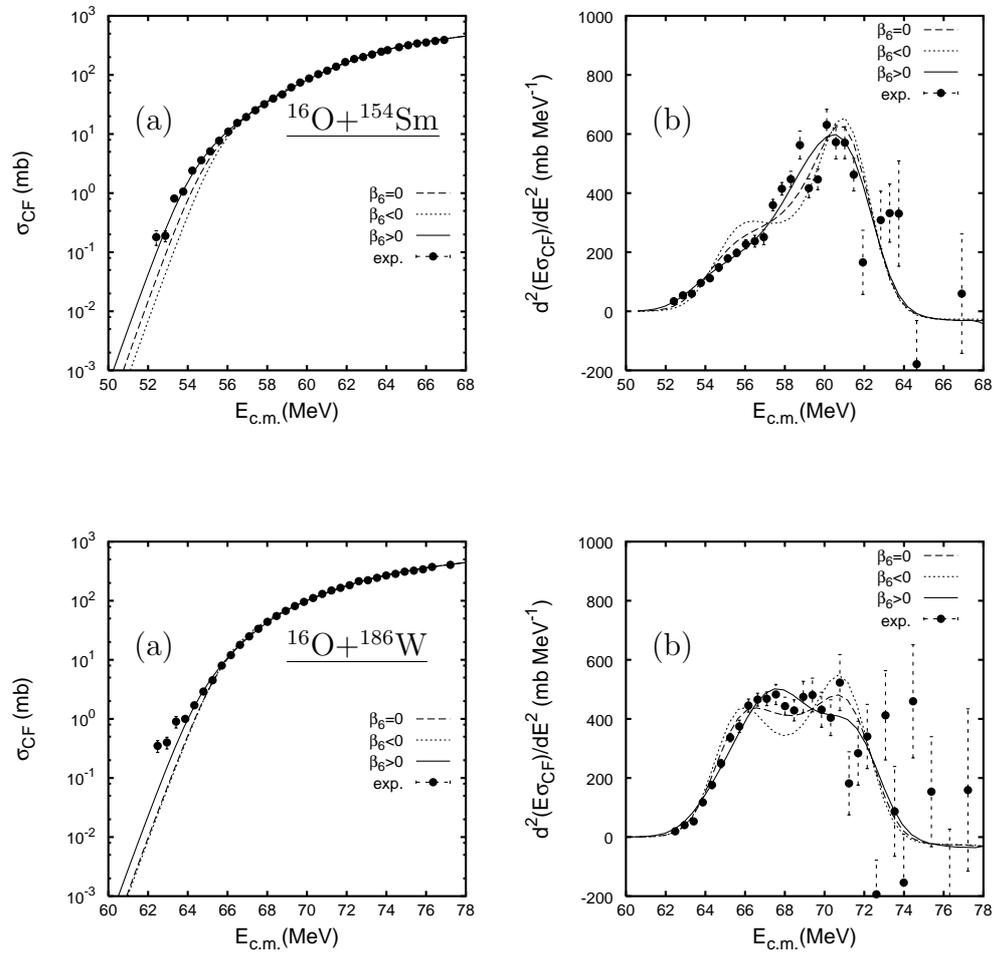
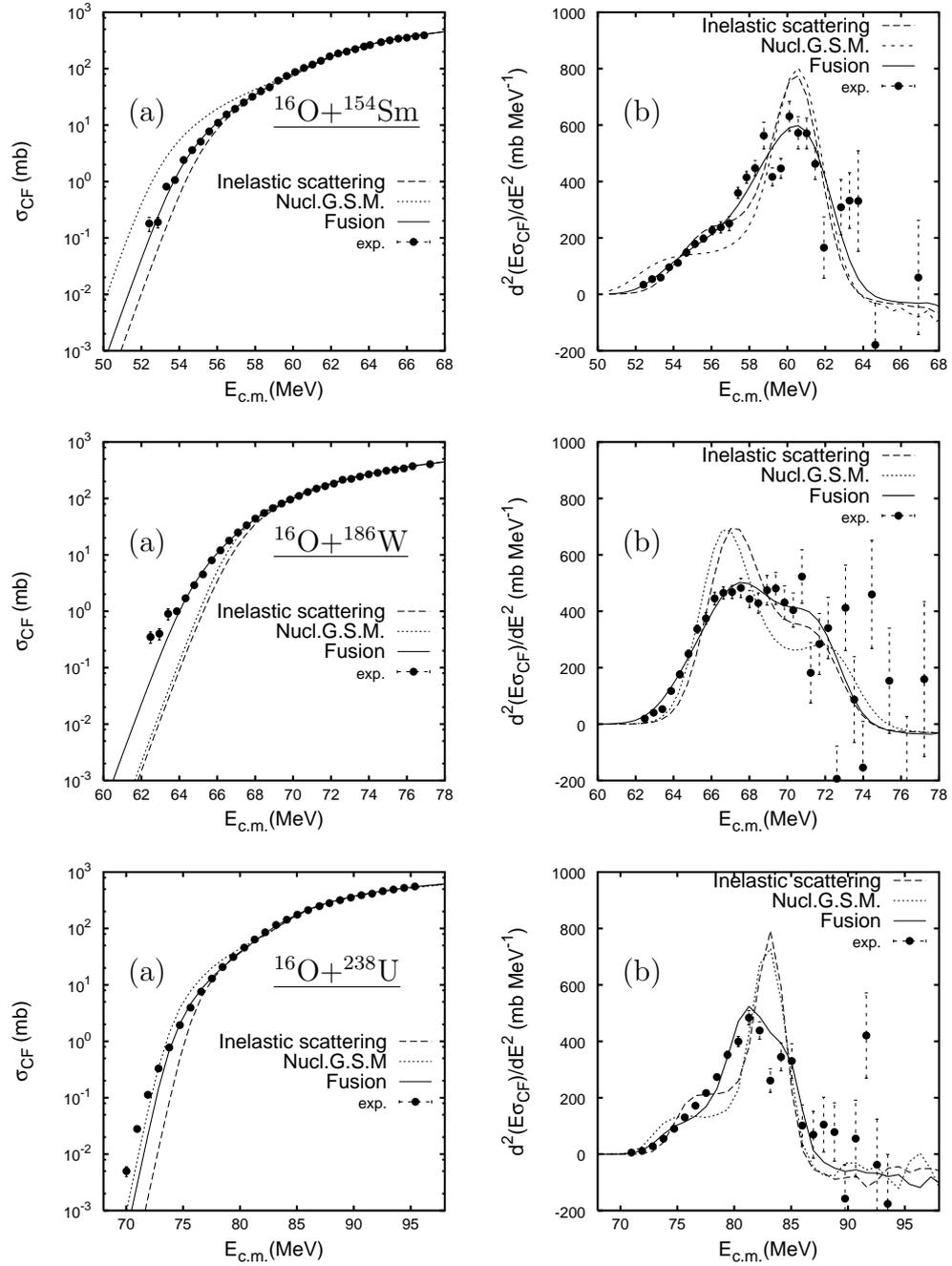
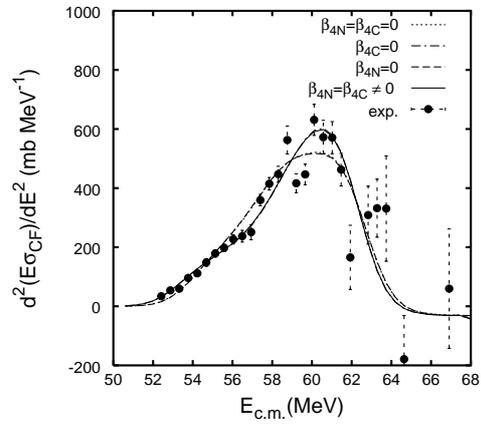
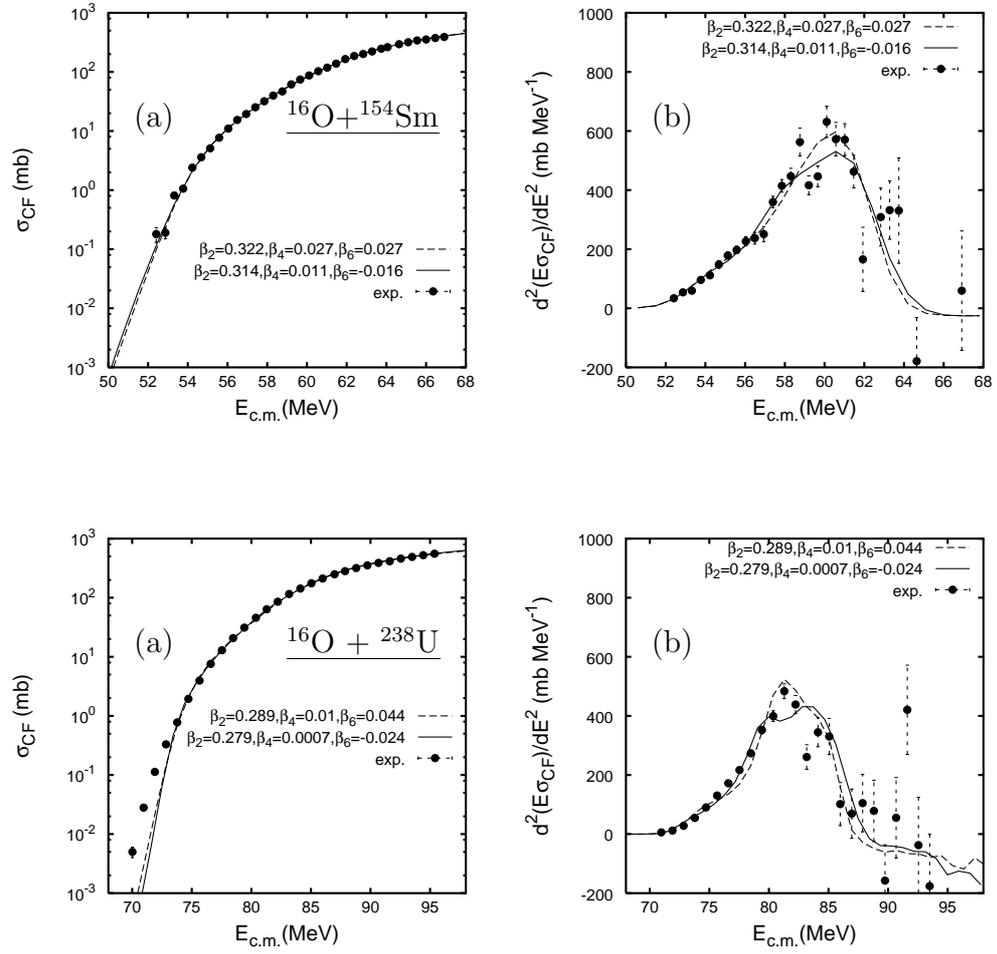
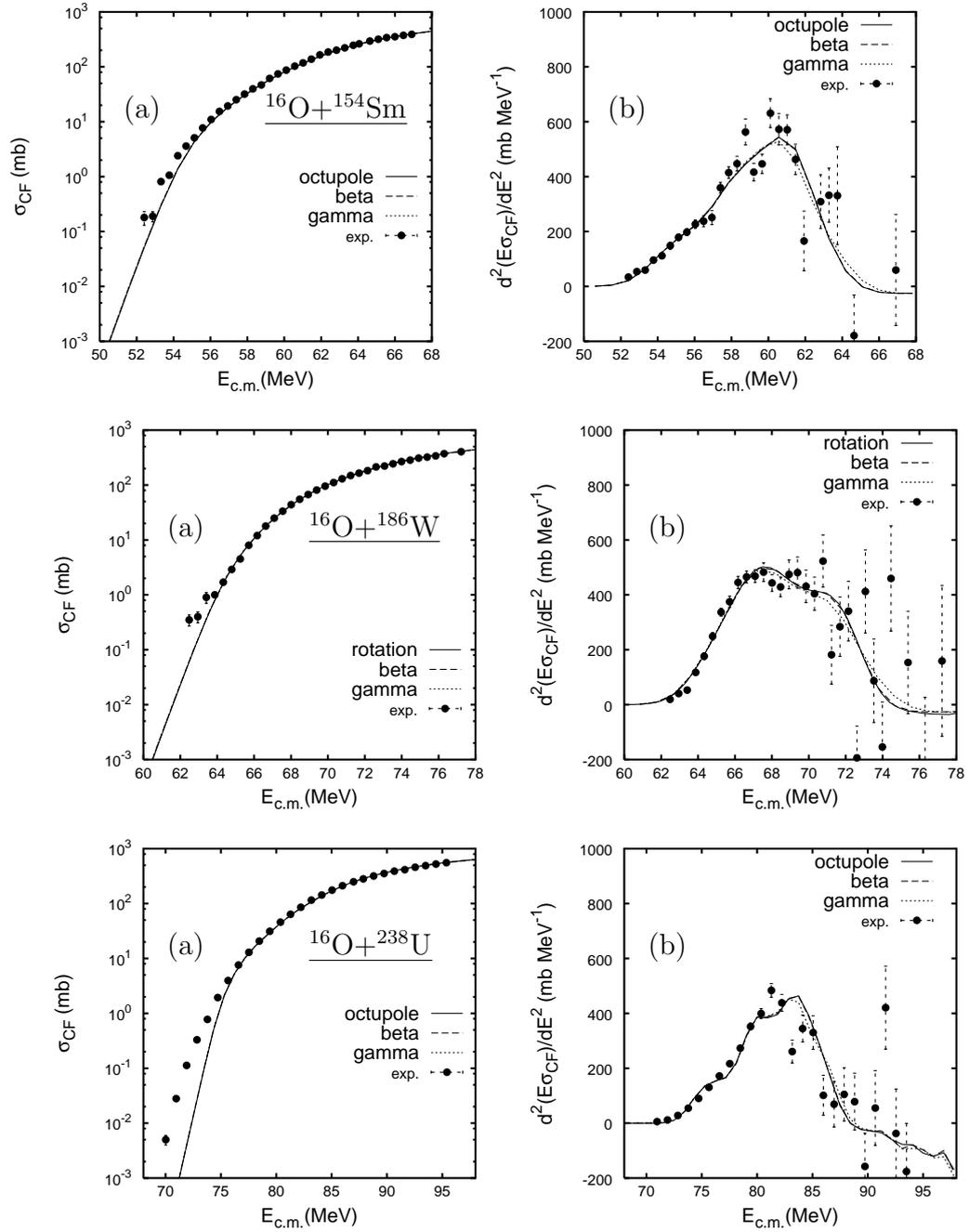
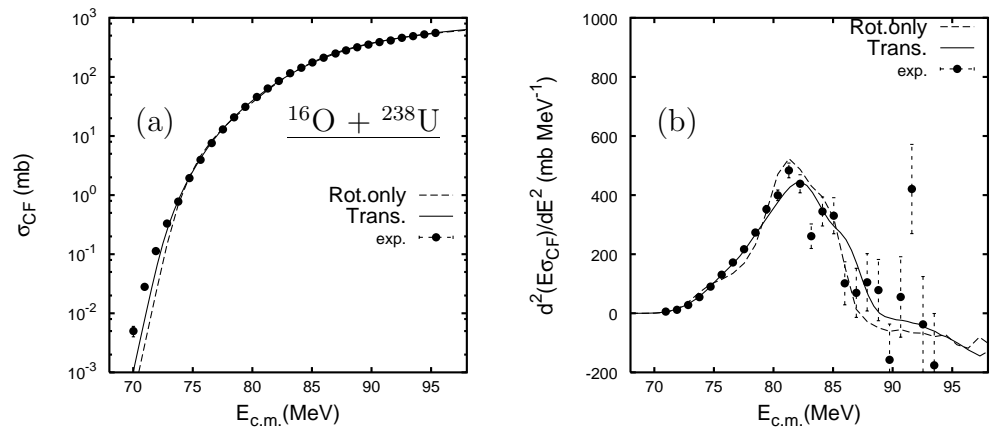
\end{document}